\documentclass[twocolumn,aps,prl,preprintnumbers]{revtex4}
\usepackage{}
\usepackage{bbm}
\usepackage[latin9]{inputenc}
\usepackage{amsmath}
\usepackage{amssymb}
\usepackage{graphicx}
\usepackage{mathrsfs}
\usepackage{amsfonts}
\usepackage{amsthm}
\usepackage{color}
\usepackage{txfonts}
\usepackage{lipsum}
\usepackage[colorlinks=true,citecolor=blue,linkcolor=blue,urlcolor=blue,anchorcolor=blue]{hyperref}%
\hypersetup{colorlinks=true,citecolor=blue,linkcolor=blue,urlcolor=blue}
\providecommand{\U}[1]{\protect\rule{.1in}{.1in}}
\makeatletter
\@ifundefined{textcolor}{}
{
\definecolor{BLACK}{gray}{0}
\definecolor{WHITE}{gray}{1}
\definecolor{RED}{rgb}{1,0,0}
\definecolor{GREEN}{rgb}{0,1,0}
\definecolor{BLUE}{rgb}{0,0,1}
\definecolor{CYAN}{cmyk}{1,0,0,0}
\definecolor{MAGENTA}{cmyk}{0,1,0,0}
\definecolor{YELLOW}{cmyk}{0,0,1,0}
}
\makeatother
\begin{document}
\title{Third-order topological insulator in three-dimensional lattice of magnetic vortices}
\author{Z.-X. Li}
\author{Zhenyu Wang}
\author{Zhizhi Zhang}
\author{Yunshan Cao}
\author{Peng Yan}
\email[Corresponding author: ]{yan@uestc.edu.cn}
\affiliation{School of Electronic Science and Engineering and State Key Laboratory of Electronic Thin Films and Integrated Devices, University of Electronic Science and Technology of China, Chengdu 610054, China}

\begin{abstract}
Recent acoustic and electrical-circuit experiments have reported the third-order (or octupole) topological insulating phase, while its counterpart in classical magnetic systems is yet to be realized. Here we explore the collective dynamics of magnetic vortices in three-dimensional breathing cuboids, and find that the vortex lattice can support zero-dimensional corner states, one-dimensional hinge states, two-dimensional surface states, and three-dimensional bulk states, when the ratio of alternating intralayer and interlayer bond lengths goes beyond a critical value. We show that only the corner states are stable against external frustrations because of the topological protection. Full micromagnetic simulations verify our theoretical predictions with good agreement. 
\end{abstract}

\maketitle
\section{INTRODUCTION}
The recent discovery of higher-order topological insulators (HOTIs) \cite{BenalcazarS2017,BenalcazarPRB2017,EzawaPRL2018,SongPRL2017,LangbehnPRL2017,SchindlerSA2018,NohNP2018,HassanNP2018,MittalNP2018,ChenPRL2019,XiePRL2019,LiNP2020,XueNM2019,NiNM2019,QiPRL2020,ZhangAM2019,ZhangNP2019,FanPRL2019,ImhofNP2018,SerraPRB2019,YangPRR2020,SongNL2020} has extensively broadened our understanding of topological phases of matter. The peculiar hinge and corner states emerging in HOTIs are attracting a lot of attention for both the fundamental interest (e.g., bulk-boundary correspondence) and the potential application in topological devices. In an $n$-dimensional system, the conventional topological insulator (TI), dubbed as first-order topological insulator (FOTI), supports $(n-1)$-dimensional topological edge states \cite{HasanRMP2010,QiRMP2011}, while the HOTI allows $(n-k)$-dimensional ($2\leq k\leq n$) topological bound states \cite{XuePRL2019,XueNC2020,NiNC2020,BaoPRB2019,ZhangNC2019,WeinerSA2020,ZhangPRB2020,ChenPRX2021}. Presently, most of HOTI states reported in the literature belong to the second-order phase ($k=2$), with a few exceptions of third-order TIs ($k=3$) being realized in acoustic and electric-circuit experiments. The main difficulty lies in the fabrication and detection of the rather spatially localized corner state in three-dimensional devices, while the artificial crystal can well overcome this issue.

Recently, the topological states in classical magnetic systems has received much interest by the spintronics community. Both the first- \cite{ZhangPRB2013,ShindouPRB2013,MookPRB2014,ChisnellPRL2015,WangPRB2017,RuckriegelPRB2018,SuPRB2017_1,SuPRB2017_2} and second-order \cite{SilJPCM2020,HirosawaPRL2020} topological phases of spin wave, one of the elementary excitations in ordered magnets, have been studied. In addition, topological insulating states emerging in metamaterials based on magnetic solitons (such as domain wall, vortex, and skyrmion) \cite{WachowiakS2002,MakhfudzPRL2012,BinzS2009,JiangS2015,CatalanRMP2012} are attracting growing interest as well. Kim \emph{et al.} \cite{KimPRL2017} and Li \emph{et al.} \cite{LiPRB2018} show that the two-dimensional honeycomb lattice of magnetic solitons can support robust chiral edge states. The realization of Su-Schrieffer-Heeger (SSH) \cite{SuPRL1979} states in one-dimensional magnetic soliton lattice is demonstrated very recently by Li \emph{et al.} \cite{LiAR2020_2} and Go \emph{et al.} \cite{GoPRB2020}. Moreover, it has been predicted that the second-order TI phases can appear in breathing kagome \cite{Linpj2019}, honeycomb \cite{LiPRA2020}, and square \cite{LiPRB2020} lattices of magnetic vortices under proper conditions. A thorough review of topological insulators and semimetals in classical magnetic systems can be found in Ref.  \cite{LiPR2021}. So far, all TI states reported in classical magnetic systems belong to the first- or second-order phase. The observation of the third-order phase is still lacking. 
\begin{figure}[ptbh]
\begin{centering}
\includegraphics[width=0.48\textwidth]{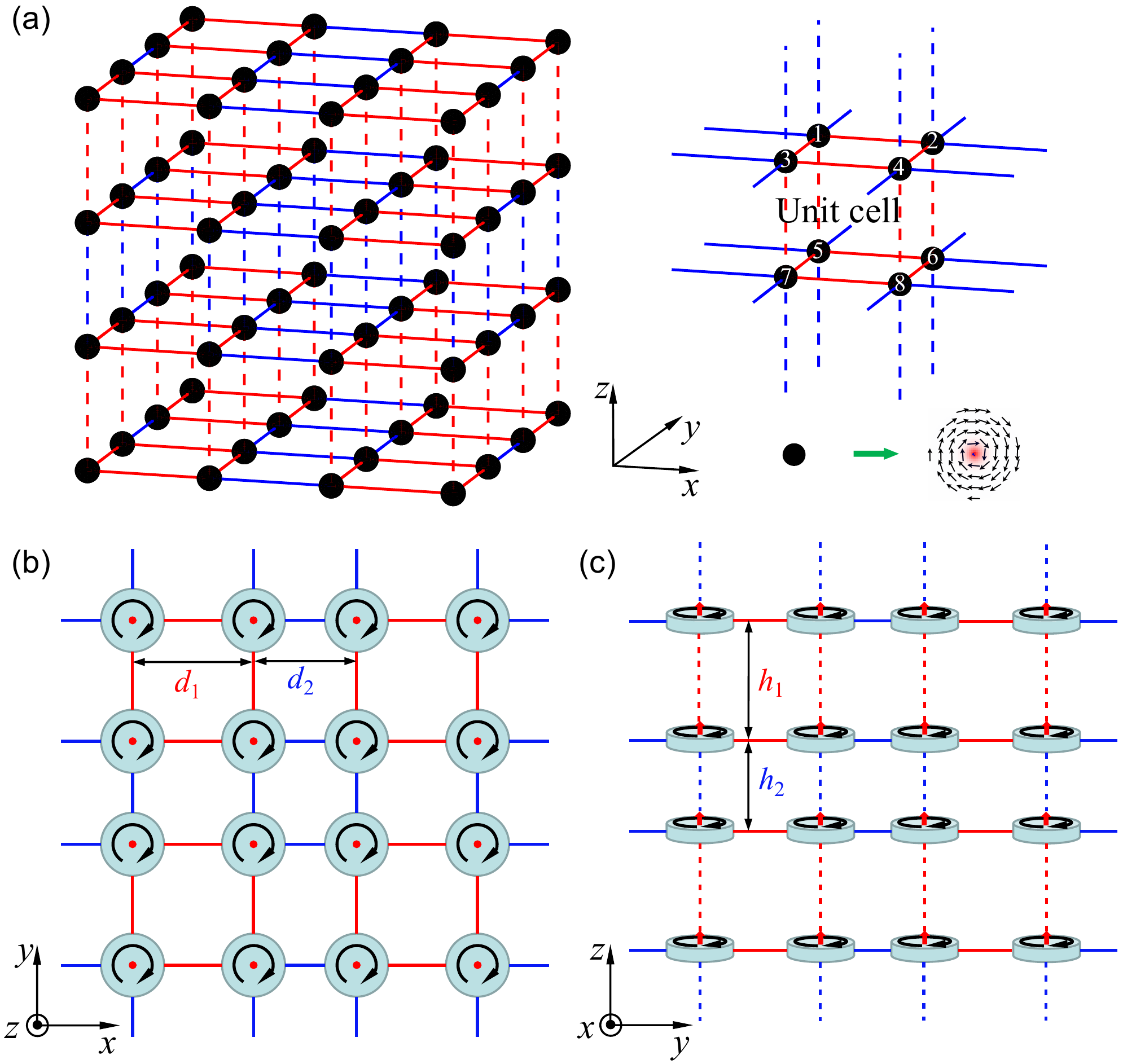}
\par\end{centering}
\caption{(a) Illustration of a three-dimensional magnetic vortex lattice, with black balls denoting nanodisks with vortex state and solid (dashed) red and blue segments representing alternating lengths of intralayer (interlayer) intracellular and intercellular bonds, respectively. Top (b) and side (c) view of the crystal structure. $d_{1}$, $d_{2}$, $h_{1}$, and $h_{2}$ are the bond lengths.}
\label{Figure1}
\end{figure}
 
In this paper, we present both analytical and numerical studies of the collective dynamics of magnetic vortices arranged in a three-dimensional breathing cuboid lattice [see Fig. \ref{Figure1}(a)]. By solving the equations of motion of interacting vortices, we obtain the band structures and predict that the third-order topological in-gap edge states (corner states) emerge when the geometric conditions $d_{1}/d_{2}>1$ and $h_{1}/h_{2}>1$ are satisfied simultaneously. Here $d_{1}$ and $d_{2}$ ($h_{1}$ and $h_{2}$) are the alternating lengths of intralayer (interlayer) intracellular and intercellular bonds, respectively, as shown in Figs. \ref{Figure1}(b) and \ref{Figure1}(c). In such condition, the one-dimensional hinge states, two-dimensional surface states, and three-dimensional bulk states are identified to be topologically trivial. These results can be understood in terms of the picture provided by the generalized SSH model. The robustness of the corner states is investigated by introducing moderate disorder and defects to this three-dimensional soliton system. We perform full micromagnetic simulations to verify theoretical predictions and find a good agreement between them. It is noted that the fabrication \cite{GuangNC2020,HanzeSP2016} and orbits tracking \cite{MollerCP2020} of magnetic soliton lattice are all within the reach of current technology. Our findings open a new route toward realizing third-order TIs in classical magnetic systems that may inspire the design of robust spintronic devices in the future. 

The outline of this paper is as follows: In Sec. \ref{section2}, we present the model and method. Topological phases of the three-dimensional vortex lattice are discussed in Sec. \ref{section3}, including theoretical calculations and micromagnetic simulations. Conclusion and outlook are drawn in Sec. \ref{section4}. 
\begin{figure*}[ptbh]
\begin{centering}
\includegraphics[width=0.9\textwidth]{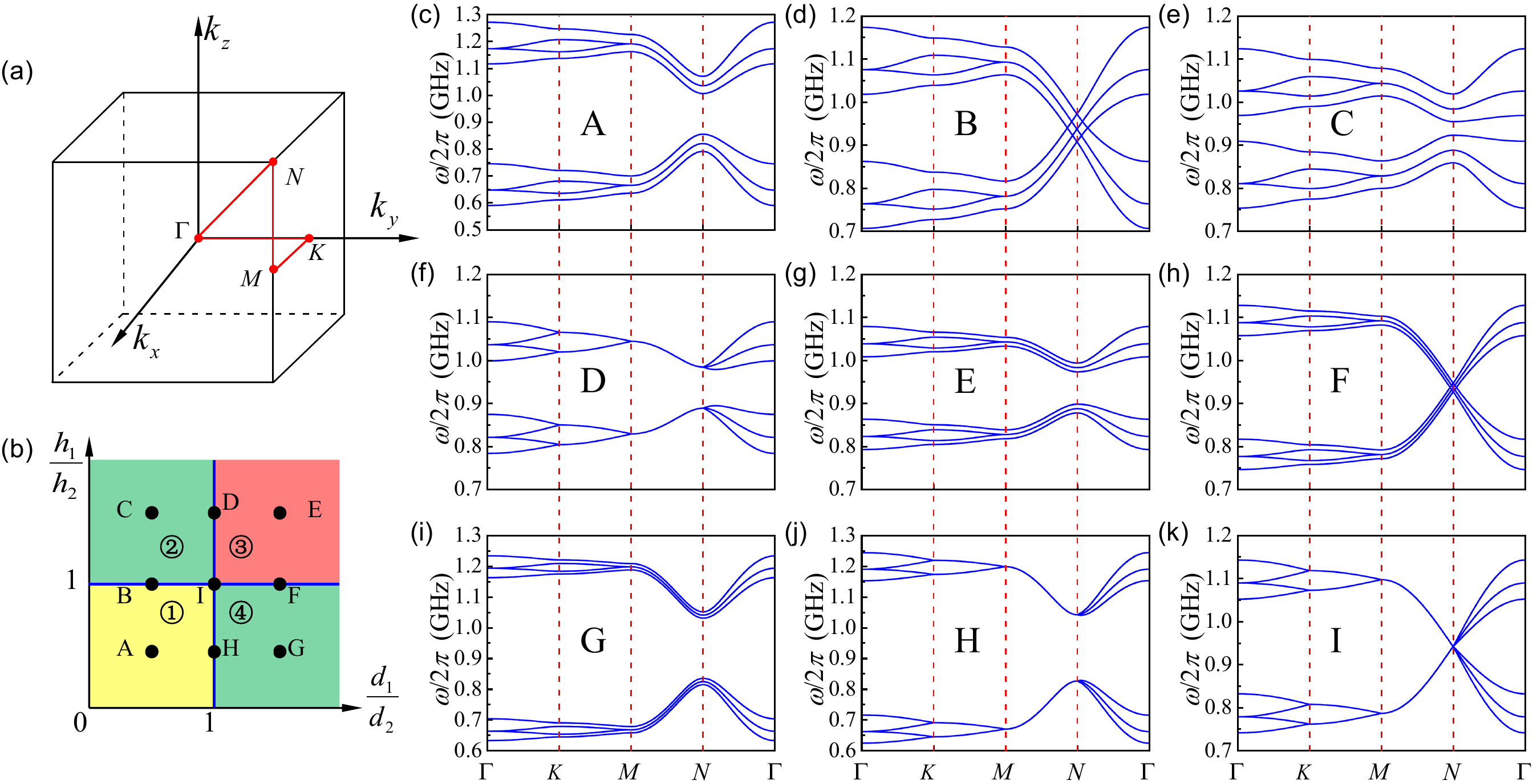}
\par\end{centering}
\caption{(a) The first Brillouin zone, with the high-symmetry points $\Gamma$, $K$, $M$, and $N$ locating at $(k_{x},k_{y},k_{z})=(0, 0, 0)$, $(0,\frac{\pi}{d_{1}+d_{2}},0)$, $(\frac{\pi}{d_{1}+d_{2}},\frac{\pi}{d_{1}+d_{2}},0)$,  and $(\frac{\pi}{d_{1}+d_{2}}, \frac{\pi}{d_{1}+d_{2}},\frac{\pi}{h_{1}+h_{2}})$, respectively. (b) The phase diagram of the system, with $d_{2}$ and $h_{2}$ being fixed to 150 nm and 70 nm, respectively. (c)-(k) The band structures along the path $\Gamma-K-M-N-\Gamma$ for different geometric parameters ($d_{1}$ and $h_{1}$) as marked by black dots in (b).}
\label{Figure2}
\end{figure*}
\section{MODEL AND METHOD}\label{section2}
A three-dimensional breathing cuboid lattice of magnetic nanodisks with vortex states is considered, as shown in Fig. \ref{Figure1}. For simplicity, we focus on the collective dynamics around the frequency of gyrotropic vortex mode and ignore both the inertial and non-Newtonian effects, which can be modeled by the Thiele's equation \cite{KimPRL2017,ThielePRL1973}:
\begin{equation}\label{Eq1}
\mathcal {G}\hat{z}\times \frac{d\textbf{U}_{j}}{dt}+\textbf{F}_{j}=0,
\end{equation}
where $\mathbf{U}_{j}= \mathbf R_{j} - \mathbf R_{j}^{0}$ is the displacement of the $j$-th vortex core from the equilibrium position $\mathbf R_{j}^{0}$, $\mathcal {G}= -4\pi$$Qw M_{s}$/$\gamma$ is the gyroscopic constant with $Q=\frac{1}{4\pi}\int \!\!\! \int{dxdy\mathbf{m}\cdot(\frac {\partial \mathbf{m}}{\partial {x} } \times \frac {\partial \mathbf{m}}{\partial y } )}$ the topological charge of the vortex state, $\mathbf {m}$ the unit vector along the local magnetization direction, $w$ the thickness of the nanodisk, $M_{s}$ the saturation magnetization, and $\gamma$ being the gyromagnetic ratio. $\textbf{F}_{j}=-\partial {\mathcal{W}} / \partial \mathbf U_{j}$ is the conservative force, where ${\mathcal{W}}$ is the total energy including both the confining potential on the vortex due to the nanodisk boundary and the (intralayer and interlayer) coupling between nearest neighbor nanodisks: $\mathcal{W}=\sum_{j}\mathcal {K}\textbf{U}_{j}^{2}/2+\sum_{j\neq k}U_{jk}/2$ with $U_{jk}=\mathcal {I}_{\parallel}U_{j}^{\parallel}U_{k}^{\parallel}-\mathcal {I}_{\perp}U_{j}^{\perp}U_{k}^{\perp}+\mu\textbf{U}_{j}\cdot\textbf{U}_{k}$ \cite{ShibataPRB2003,ShibataPRB2004,GuslienkoAPL2005}. Here, $\mathcal {K}$ is the spring constant, $\mathcal {I}_{\parallel}$ ($\mathcal {I}_{\perp}$) is the intralayer longitudinal (transverse) coupling constant, and $\mu$ is the interlayer coupling parameter. 

Imposing $\textbf{U}_{j}=(u_{j},v_{j})$ and defining $\psi_{j}=u_{j}+iv_{j}$, Eq. \eqref{Eq1} can be re-written as:  
\begin{widetext}
\begin{equation}\label{Eq2}
  \begin{aligned}
-i\dot{\psi}_{j}=(\omega_{0}-\frac{\xi^{2}_{1}+\xi^{2}_{2}}{\omega_{0}})\psi_{j}+\sum_{k\in\langle j\rangle,l}\zeta_{l}\psi_{k}+\sum_{k\in\langle j'\rangle,n}\eta_{n}\psi_{k}-\frac{\xi_{1}\xi_{2}}{2\omega_{0}}\sum_{s\in\langle\langle j_{1}\rangle\rangle}e^{i2\bar{\theta}_{js}}\psi_{s}-\frac{\xi^{2}_{2}}{2\omega_{0}}\sum_{s\in\langle\langle j_{2}\rangle\rangle}e^{i2\bar{\theta}_{js}}\psi_{s}-\frac{\xi^{2}_{1}}{2\omega_{0}}\sum_{s\in\langle\langle j_{3}\rangle\rangle}e^{i2\bar{\theta}_{js}}\psi_{s},
  \end{aligned}
\end{equation}
\end{widetext}
where ${\omega}_{0}=\mathcal{K}/|\mathcal{G}|$, $\zeta_{l}=(\mathcal {I}_{\parallel, l}-\mathcal {I}_{\perp, l})/2\mathcal {|G|} $, and $\xi_{l}=(\mathcal {I}_{\parallel, l}+\mathcal {I}_{\perp, l})/2\mathcal {|G|}$, in which $l=1$ ($l=2$) represents the intralayer intracellular (intercellular) connection; $\eta_{n}=\mu_{n}/|\mathcal{G}|$, $n=1$ ($n=2$) denotes the interlayer intracellular (intercellular) bond; $\bar{\theta}_{js}=\theta_{jk}-\theta_{ks}$ is the relative angle from the bond $k\rightarrow s$ to the bond $j\rightarrow k$ with $k$ between $j$ and $s$, and $\langle j\rangle$ and $\langle j'\rangle$ are the set of intralayer and interlayer nearest neighbors of $j$, respectively; $\langle\langle j_{1}\rangle\rangle$, $\langle\langle j_{2}\rangle\rangle$, and $\langle\langle j_{3}\rangle\rangle$ represent the intralayer next-nearest neighbors of $j$.

To solve Eq. \eqref{Eq2} numerically, the key parameters $\mathcal{K}$, $\mathcal {I}_{\parallel}$, $\mathcal {I}_{\perp}$, and $\mu$ should be determined. Firstly, the spring constant $\mathcal{K}$ can be obtained from the relation $\mathcal{K}=\omega_{0}|\mathcal{G}|$, with $\omega_{0}$ the gyrotropic frequency of a single vortex. For Permalloy (Py) \cite{YooAPL2012,VeltenAPL2017} nanodisk of vortex state with thickness $w=10$ nm and radius $r=50$ nm, the gyrotropic frequency $\omega_{0}=2\pi\times0.939$ GHz, gyroscopic constant $\mathcal{G}=-3.0725\times10^{-13}$ J\,s\,rad$^{-1}$m$^{-2}$ ($Q=1/2$) \cite{Linpj2019}, we have $\mathcal{K}=1.8128\times10^{-3}$ J\,m$^{-2}$. Secondly, the analytical expressions of $\mathcal {I}_{\parallel}$ and $\mathcal {I}_{\perp}$ on the distance $d$ between vortices have been obtained in a simplified two-nanodisk system \cite{Linpj2019}. Finally, the dependence of parameter $\mu$ on the distance $h$ can also be determined from micromagnetic simulations by considering a stacking two-vortex system (see Appendix A for details).  

For an infinite three-dimensional vortex lattice, the unit cell can be selected as shown in Fig. \ref{Figure1}(a). The three basis vectors of the system are $\textbf{a}_{1}=(d_{1}+d_{2})\hat{x}$, $\textbf{a}_{2}=(d_{1}+d_{2})\hat{y}$, and $\textbf{a}_{3}=(h_{1}+h_{2})\hat{z}$, respectively. By performing a plane wave expansion of $\psi_{j}$, we obtain the matrix form of the Hamiltonian expressed in the momentum space: 
\begin{equation}\label{Eq3}
 \mathcal {H}=\left(
 \begin{matrix}
   Q_{0} & Q_{1} & Q_{2} & Q_{3} & Q_{4} & 0 & 0 & 0 \\
  Q_{1}^{*} & Q_{0} & Q_{5} & Q_{2} & 0 & Q_{4} & 0 & 0 \\
  Q_{2}^{*} & Q_{5}^{*} & Q_{0}& Q_{1} & 0 & 0 & Q_{4} & 0\\
   Q_{3}^{*} & Q_{2}^{*}& Q_{1}^{*} & Q_{0} & 0 & 0 & 0 & Q_{4}\\
Q_{4}^{*} & 0 & 0 & 0 & Q_{0} & Q_{1} & Q_{2} & Q_{3}\\
0 & Q_{4}^{*} & 0 & 0 & Q_{1}^{*} & Q_{0} & Q_{5} & Q_{2}\\
0 & 0 & Q_{4}^{*} & 0 & Q_{2}^{*} & Q_{5}^{*} & Q_{0}& Q_{1}\\
0 & 0 & 0 & Q_{4}^{*} & Q_{3}^{*} & Q_{2}^{*}& Q_{1}^{*} & Q_{0}\\
  \end{matrix}
  \right),
\end{equation}
with the matrix elements:
\begin{widetext}
\begin{equation}\label{Eq4}
\begin{aligned}
Q_{0}&=\omega_{0}-\frac{\xi_{1}^{2}+\xi_{2}^{2}}{{\omega}_{0}}-\frac{\xi_{1}\xi_{2}}{{\omega}_{0}}[\cos(\textbf{k}\cdot\textbf{a}_{1})+\cos(\textbf{k}\cdot\textbf{a}_{2})], \\
Q_{1}&=\zeta_{2}+\zeta_{1}\exp(-i\textbf{k}\cdot\textbf{a}_{1}),\\
Q_{2}&=\zeta_{2}+\zeta_{1}\exp(i\textbf{k}\cdot\textbf{a}_{2}),\\
Q_{3}&=\frac{\xi_{1}\xi_{2}}{{\omega}_{0}}[\exp(i\textbf{k}\cdot\textbf{a}_{2})+\exp(-i\textbf{k}\cdot\textbf{a}_{1})]+\frac{\xi^{2}_{1}}{{\omega}_{0}}\exp[i\textbf{k}\cdot(\textbf{a}_{2}-\textbf{a}_{1})]+\frac{\xi^{2}_{2}}{{\omega}_{0}}, \\
Q_{4}&=\eta_{1}+\eta_{2}\exp(i\textbf{k}\cdot\textbf{a}_{3}),\\
Q_{5}&=\frac{\xi_{1}\xi_{2}}{{\omega}_{0}}[\exp(i\textbf{k}\cdot\textbf{a}_{2})+\exp(i\textbf{k}\cdot\textbf{a}_{1})]+\frac{\xi^{2}_{1}}{{\omega}_{0}}\exp[i\textbf{k}\cdot(\textbf{a}_{2}+\textbf{a}_{1})]+\frac{\xi^{2}_{2}}{{\omega}_{0}},
\end{aligned}
\end{equation}
\end{widetext}where $\mathbf{k}$ is the wave vector.

The spectrum of Hamiltonian \eqref{Eq3} fully determines the phase allowed in a finite crystal due to the principle of bulk-boundary correspondence. In what follows, we present detailed calculations and discussions.
\section{Third-order topological insulator}\label{section3}
\subsection{Theoretical calculations}
\begin{figure*}[ptbh]
\begin{centering}
\includegraphics[width=0.9\textwidth]{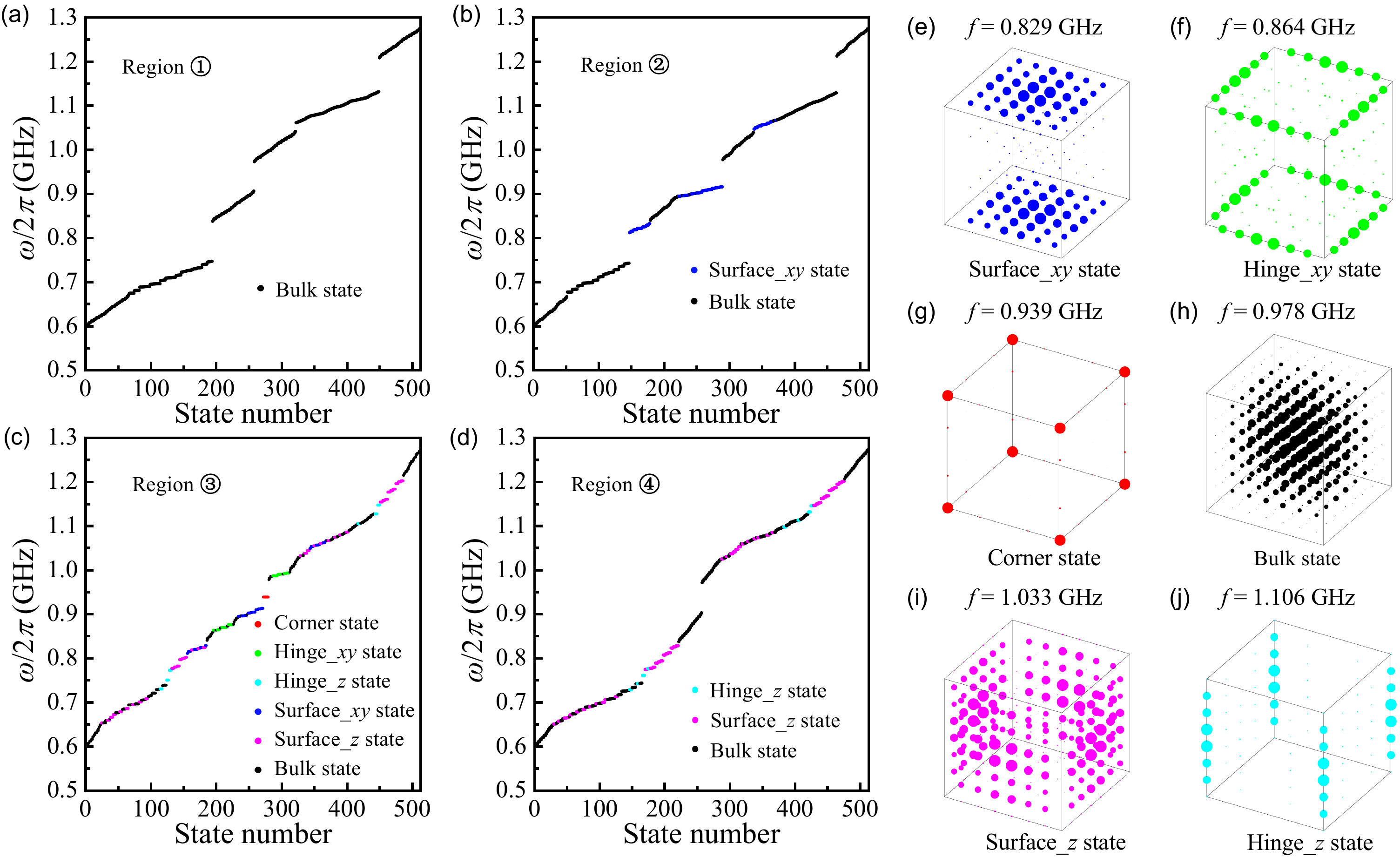}
\par\end{centering}
\caption{Eigenfrequencies of the collective vortex gyration in a three-dimensional $8\times 8 \times 8$ system for parametric region \textcircled{\scriptsize{1}}  (a), parametric region \textcircled{\scriptsize{2}} (b), parametric region \textcircled{\scriptsize{3}} (c), and parametric region \textcircled{\scriptsize{4}} (d), as marked in Fig. \ref{Figure2}(b). (e)-(j) The spatial distribution of different modes. Here the size of the colorful balls is proportional to the amplitude of the wavefunction.}
\label{Figure3}
\end{figure*}
Firstly, we calculate the bulk band structure along the path $\Gamma-K-M-N-\Gamma$ in the first Brillouin zone [see Fig. \ref{Figure2}(a)] for different geometric parameters, as shown in Figs. \ref{Figure2}(c)-\ref{Figure2}(k), with the parameters $d_{2}$ and $h_{2}$ being fixed to 150 nm and 70 nm, respectively. One can see that the system supports eight bands due to the interlayer coupling between nanodisks, in contrast to the four-band structure in a two-dimensional breathing square vortex lattice obtained in Ref. \cite{LiPRB2020}. For parameters $(d_{1},h_{1})=$ (150 nm, 70 nm), as marked by letter ``I" in Fig. \ref{Figure2}(b), we can see that all bands merge together, leading to a gapless band structure, as shown in Fig. \ref{Figure2}(k). On the one hand, if we keep $d_{1}/d_{2}=1$ and let $h_{1}/h_{2}\neq1$, a gap will open at the high-symmetry $N$ point [see Figs. \ref{Figure2}(f) and \ref{Figure2}(j)], where the letters ``D" and ``H" represent parameters $(d_{1}, h_{1})=$ (150 nm, 110 nm) and (150 nm, 40 nm), respectively. On the other hand, if we keep $h_{1}/h_{2}=1$, while let $d_{1}/d_{2}\neq1$, a gap opens at the high-symmetry $K$ and $M$ points [see Figs. \ref{Figure2}(d) and \ref{Figure2}(h)], with the letters ``B" and ``F" representing parameters $(d_{1}, h_{1})=$ (120 nm, 70 nm) and (180 nm, 70 nm), respectively. Furthermore, if we set $d_{1}/d_{2}\neq1$ and $h_{1}/h_{2}\neq1$ simultaneously, the gap opens at all high-symmetry points, as shown in Figs. \ref{Figure2}(c), \ref{Figure2}(e), \ref{Figure2}(g), and \ref{Figure2}(i), with the parameters $(d_{1}, h_{1})=$ (120 nm, 40 nm), (120 nm, 110 nm), (180 nm, 110 nm), and (180 nm, 40 nm), respectively. To further distinguish whether these insulating phases are topologically protected, one should study the eigen modes in finite systems and the robustness of them.   

To find the higher-order topological edge states, we calculate the eigenfrequencies of a finite system with 512 ($8\times 8 \times 8$) nanodisks for four different parameters regions [see Fig. \ref{Figure2}(b)]. For region \textcircled{\scriptsize{1}}, we set $(d_{1},d_{2},h_{1},h_{2})=$ (104 nm, 180 nm, 40 nm, 110 nm), with the eigenfrequencies of the system shown in Fig. \ref{Figure3}(a). By analyzing the spatial distribution of the wavefunction, we identify that the system can only support the bulk states [see Fig. \ref{Figure3}(h)]. Then we consider region \textcircled{\scriptsize{2}}, with the geometric parameters $(d_{1},d_{2},h_{1},h_{2})=$ (104 nm, 180 nm, 110 nm, 40 nm), and the eigenfrequencies are shown in Fig. \ref{Figure3}(b), from which one can see that the system can support both the surface$\_xy$ states and the bulk states, with the spatial distribution of the surface$\_xy$ states plotted in Fig. \ref{Figure3}(e). For region \textcircled{\scriptsize{3}}, we set $(d_{1},d_{2},h_{1},h_{2})=$ (180 nm, 104 nm, 110 nm, 40 nm), and the spectrum is plotted in Fig. \ref{Figure3}(c). One can clearly see that five different edge states emerge, including the zero-dimensional corner states [see Fig. \ref{Figure3}(g)], two one-dimensional hinge states [see Figs. \ref{Figure3}(f) and \ref{Figure3}(j)], and two two-dimensional surface states [see Figs. \ref{Figure3}(e) and \ref{Figure3}(i)]. Finally, for region \textcircled{\scriptsize{4}}, we set parameters $(d_{1},d_{2},h_{1},h_{2})=$ (180 nm, 104 nm, 40 nm, 110 nm). In such case, we observe the hinge$\_z$, surface$\_z$, and bulk states [see Figs. \ref{Figure3}(j), \ref{Figure3}(i), and \ref{Figure3}(h)]. 

To judge if the emerging edge states in Fig. \ref{Figure3}(c) are topologically protected, we calculate the spectrum of the system in the presence of  disorder and defects (see Appendix B for details). We find that only the corner states are topological stable, while other edge modes (hinge and surface states) are trivial and sensitive to the introduced disturbances. We therefore realize the third-order TI in classical magnetic system based on vortex metamaterials. These results can be straightforwardly understood by the SSH model. On the one hand, as shown in Ref. \cite{LiPRB2020}, we have proved that a two-dimensional breathing square lattice of vortices can support second-order topological edge state when $d_{1}/d_{2}>1$. On the other hand, the layered structure can be viewed as an one-dimensional analogue of the SSH model. Then, only when $h_{1}/h_{2}>1$, the system can support topological edge state. We thus conclude that the third-order topological insulating phase can appear if and only if the geometric conditions $d_{1}/d_{2}>1$ and $h_{1}/h_{2}>1$ are satisfied simultaneously. 
\begin{figure}[ptbh]
\begin{centering}
\includegraphics[width=0.48\textwidth]{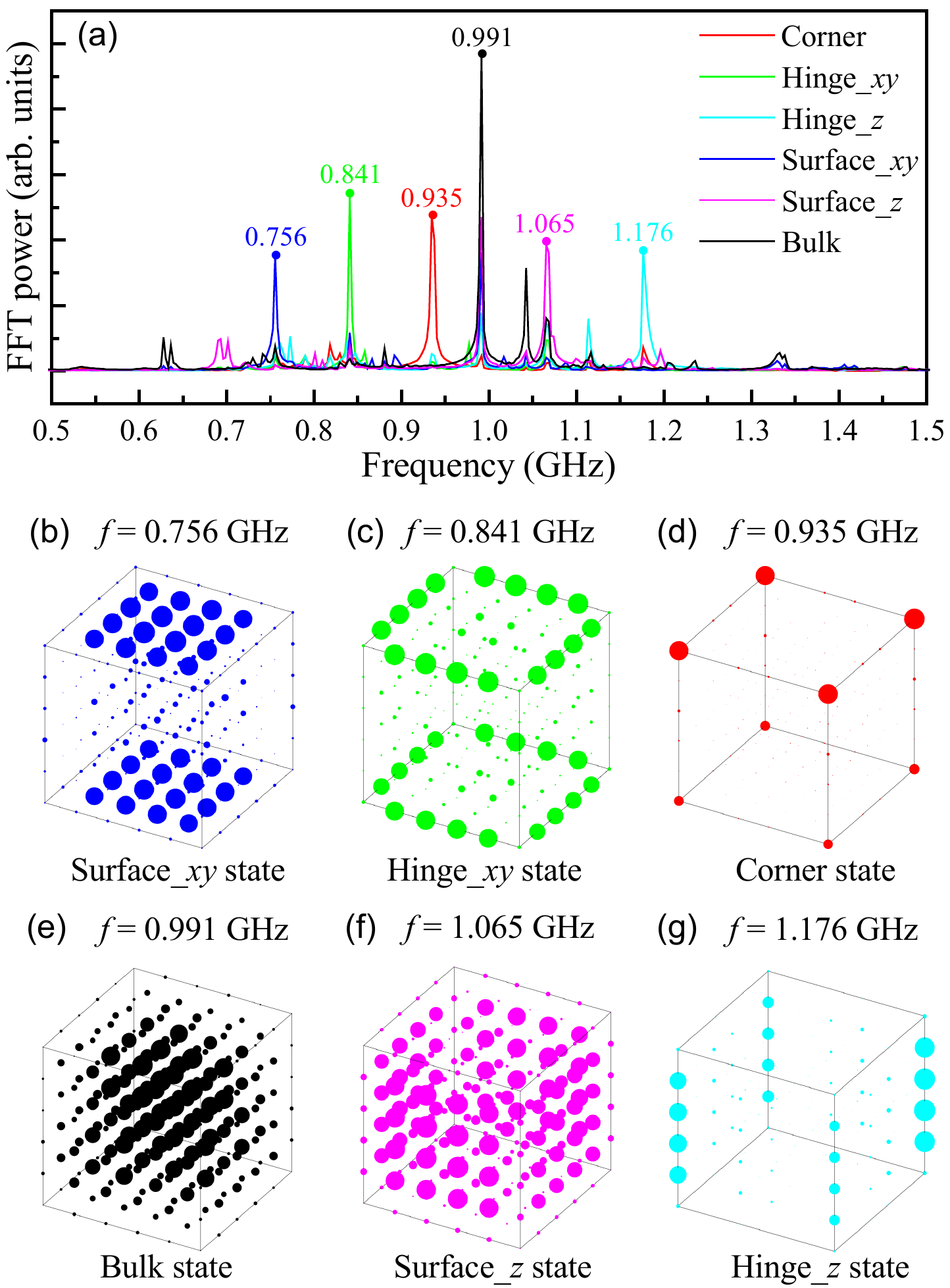}
\par\end{centering}
\caption{(a) The temporal Fourier spectra of the vortex oscillations at different positions. (b)-(g) The spatial distribution of FFT intensity with different frequencies as marked in (a). The size of the balls reflects the strength of the oscillation.}
\label{Figure4}
\end{figure}     
\subsection{Micromagnetic simulations}
To confirm the theoretical predictions above, we perform full micromagnetic simulations. All materials parameters adopted in the simulations are the same as those in the theoretical calculations presented in Fig. \ref{Figure3}(c), with the saturation magnetization $M_{s}=0.86\times10^{6}$ A/m, the exchange stiffness $A=1.3\times10^{-11}$ J/m, and the Gilbert damping constant $\alpha=10^{-4}$. To reduce the computational workload, we consider the $6\times6\times6$ nanodisks system in the simulation. The micromagnetic package MUMAX3 \cite{Vansteenkiste2014} is used to simulate the collective dynamics of the vortex lattice. Here the cellsize is set to $2\times2\times10 $ nm$^{3}$. A sinc-function magnetic field $H(t)=H_{0}\sin[2\pi$\emph{f}$(t-t_{0})]/[2\pi$\emph{f}$(t-t_{0})]$ along the $x$-direction with $H_{0}=10$ mT, $f=30$ GHz, and $t_{0}=1$ ns is applied to the whole system for 352 ns. Then the full spectrum of the system can be obtained by analyzing the collective oscillation of vortex lattice. The position of vortex cores $\textbf{R}_{j}=(R_{j,x}, R_{j,y}$) in all nanodisks are recorded every 200 ps. Here, $R_{j,x}=\frac {\int \!\!\! \int{x|m_{z}|^{2}dxdy}}{\int \!\!\! \int{|m_{z}|^{2}dxdy}}$, and $R_{j,y}=\frac {\int \!\!\! \int{y|m_{z}|^{2}dxdy}}{\int \!\!\! \int{|m_{z}|^{2}dxdy}}$, with the integral region confined in the $j$-th nanodisk.   
\begin{figure}[ptbh]
\begin{centering}
\includegraphics[width=0.48\textwidth]{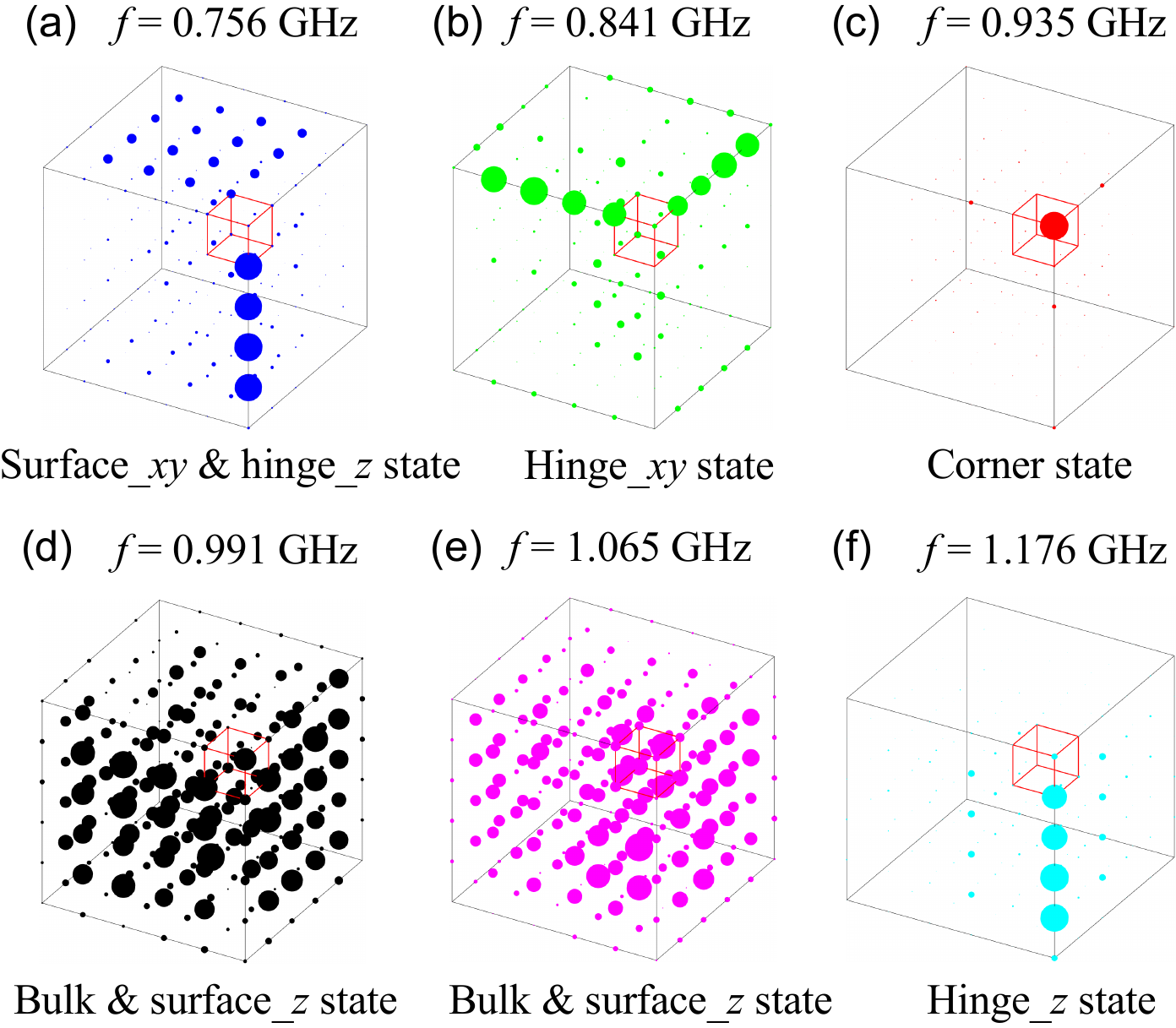}
\par\end{centering}
\caption{The spatial distribution of vortex oscillations under the local exciting field with different frequencies.}
\label{Figure5}
\end{figure}
The temporal fast Fourier transform (FFT) spectra of the vortex oscillations at different positions are plotted in Fig. \ref{Figure4}(a). Here the red, green, cyan, blue, magenta, and black curves denote the positions of corner, hinge$\_xy$, hinge$\_z$, surface$\_xy$, surface$\_z$, and bulk bands, respectively. From Fig. \ref{Figure4}(a), one can clearly identify the frequency range supporting different modes. To visualize the spatial distribution of vortex oscillation for different modes, we plot the spatial distribution of the FFT intensity with six representative frequencies [see Fig. \ref{Figure4}(a)], as shown in Figs. \ref{Figure4}(b)-\ref{Figure4}(g). Here, the radius of the balls is proportional to the FFT strength. We identify the following modes: (i) Surface$\_xy$ state with oscillation localized at top and bottom surfaces [see Fig. \ref{Figure4}(b)]; (ii) Hinge$\_xy$ state with oscillation localized at eight edges of top and bottom surfaces [see Fig. \ref{Figure4}(c)]; (iii) Corner state with oscillation localized at eight corners [see Fig. \ref{Figure4}(d)]; (iv) Bulk state with oscillation spread over the system except the surfaces [see Fig. \ref{Figure4}(e)]; (v) Surface$\_z$ state with oscillation localized at four side surfaces [see Fig. \ref{Figure4}(f)]; (vi) Hinge$\_z$ state with oscillation localized at four side edges [see Fig. \ref{Figure4}(g)]. These results agree well with theoretical calculations.         
       
In the above simulations, the external driving source (of the sinc function) is applied over the whole lattice, which prevents us from observing the time-evolution of each mode. Here, we stimulate the collective vortex dynamics by applying a sinusoidal magnetic field $\mathbf{h}(t)=h_{0}\text{sin}(2\pi f t)\hat{x}$ with $h_{0}=0.01$ mT and $t=60$ ns, localized at eight nanodisks at one corner, as denoted by red cubes in Fig. \ref{Figure5}. For different frequencies, we plot the spatial distribution of vortex oscillations, as shown in Fig. \ref{Figure5}(a)-\ref{Figure5}(f), with the size of balls representing the amplitude of vortex oscillations. One can clearly see that the spatial distribution of hinge$\_xy$ state, corner state, and hinge$\_z$ state compare well with the corresponding modes plotted in Fig. \ref{Figure4}, while other three modes are not. Interestingly, we observe various hybridized modes, including surface$\_xy$ $\&$ hinge$\_z$ state [Fig. \ref{Figure5}(a)] and bulk $\&$ surface$\_z$ state [Figs. \ref{Figure5}(d) and \ref{Figure5}(e)]. The reason for the mode hybridization is that the frequencies of these states are close and their wavefunctions have significant overlap (see Fig. \ref{Figure4}).             

\section{CONCLUSION AND OUTLOOK}\label{section4}
To conclude, we predicted the third-order TI in a three-dimensional breathing cuboid lattice of magnetic vortices. The geometric condition to observe this novel phase is analyzed by the SSH model. We showed that the third-order corner states are robust against moderate disorder and defects. Full micromagnetic simulations were implemented to verify theoretical predictions with a great agreement. Our findings pave a new way for realizing third-order TI in classical magnetic systems, which should have potential applications for designing robust spintronics devices. In this work, we have assumed that the polarity and chirality of all vortices are identical, while the effect of different polarities and chiralities on the higher-order topology is an interesting topic for future research.     

\begin{acknowledgments}
\section*{ACKNOWLEDGMENTS}
This work was supported by the National Natural Science Foundation of China (NSFC) (Grants No. 12074057, No. 11604041, and No. 11704060). Z.-X. Li acknowledges financial support from the China Postdoctoral Science Foundation (Grant No. 2019M663461) and the NSFC (Grant No. 11904048). Z. Wang was supported by the China Postdoctoral Science Foundation under Grant No. 2019M653063. Z. Zhang acknowledges the support by the China Postdoctoral Science Foundation under Grant No. 2020M673180.
\end{acknowledgments}

\section*{APPENDIX A: THE DETERMINATION OF PARAMETER}\label{APPENDIX A}
\begin{figure}[ptbh]
\begin{centering}
\includegraphics[width=0.48\textwidth]{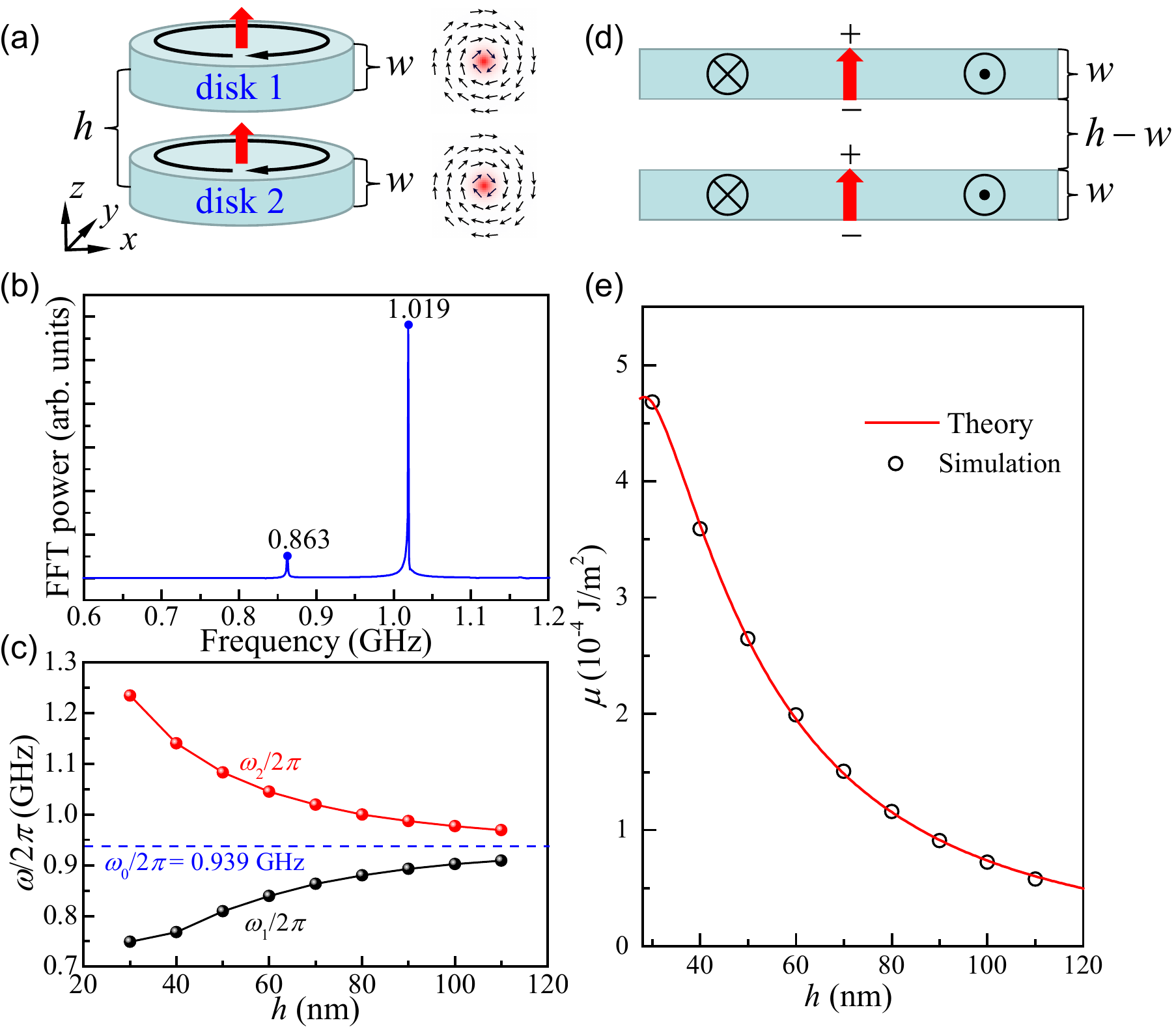}
\par\end{centering}
\caption{(a) Schematic plot of the stacking two-vortex system. The micromagnetic structures of vortices are also shown. (b) The Fourier spectrum of the vortices oscillation for $h=70$ nm. (c) The two eigen frequencies of the two-vortex system varying with $h$. (d) The illustration of the interaction between magnetic charges. (e) Dependence of the interlayer coupling parameter $\mu$ on $h$. Black circles denote simulation results and red solid line represents the analytical formula.}
\label{Figure6}
\end{figure}

The interlayer coupling parameter $\mu$ is very important for solving the dynamical equation of vortex motion. Here we determine this parameter by considering the dynamics of a two-nanodisk system with vortex state, as shown in Fig. \ref{Figure6}(a). For such case, the potential energy can be expressed as $\mathcal {W}=\sum_{j}\mathcal {K}\textbf{U}_{j}^{2}/2+\sum_{j\neq k}\mu\textbf{U}_{j}\cdot\textbf{U}_{k}/2$. By imposing $\psi_{j}=u_{j}+iv_{j}$ and $\psi_{j}=\psi_{j} e^{i\omega t}$, the dynamical equations for the two-vortex system can be simplified to:  
\begin{equation}\label{Eq5}
\begin{aligned}
\omega\psi_{1}&=\frac{\mathcal{K}}{|\mathcal{G}|}\psi_{1}+\frac{\mu}{|\mathcal{G}|}\psi_{2},\\
\omega\psi_{2}&=\frac{\mathcal{K}}{|\mathcal{G}|}\psi_{2}+\frac{\mu}{|\mathcal{G}|}\psi_{1}.
\end{aligned}
\end{equation}

By solving Eq. \eqref{Eq5}, we can obtain two resonant frequencies: $\omega_{1}=(\mathcal{K}-\mu)/|\mathcal{G}|$ and $\omega_{2}=(\mathcal{K}+\mu)/|\mathcal{G}|$. Therefore, the interlayer coupling parameter $\mu$ can be expressed as $\mu=(\omega_{2}-\omega_{1})|\mathcal{G}|/2$. Figure \ref{Figure6}(b) plots the Fourier spectrum of the vortex oscillation for $h=70$ nm. The frequencies of the peaks read $\omega_{1}/2\pi=0.863$ GHz and $\omega_{2}/2\pi=1.019$ GHz, then we obtain $\mu=1.5058\times10^{-4}$ J\,m$^{-2}$. Similarly, we can get the frequencies of the two-vortex system by varying $h$. Figure \ref{Figure6}(c) plots the $h$-dependence of $\omega_{1}$ and $\omega_{2}$. It shows that, with the increasing of $h$, $\omega_{1}$ increases while $\omega_{2}$ decreases. One naturally expects $\omega_{1}=\omega_{2}=\omega_{0}$ when $d\rightarrow \infty$. The dipole interaction between two vortices can be simplified as the interaction between magnetic charges of the vortex cores \cite{CherepovPRL2012,BondarenkoPRB2019}, as illustrated in Fig. \ref{Figure6}(d). Further, the interaction between magnetic charges can be divide into three parts: (i) The attractive interaction between negative magnetic charge of disk 1 and positive magnetic charge of disk 2 [$\propto1/(h-w)$];  (ii) The repulsive interaction of the same magnetic charge between disk 1 and disk 2 ($\propto1/h$); (iii) The attractive interaction between positive magnetic charge of disk 1 and negative magnetic charge of disk 2 [$\propto1/(h+w)$]. To obtain the analytical expression of $\mu$ on $h$, we use $c_{1}/(h-w)+c_{2}/h+c_{3}/(h+w)$ to fit the simulation results. The curve in Fig. \ref{Figure6}(e) shows the best fit of the numerical date with $c_{1}=-5.7478\times10^{-11}$ J\,m$^{-1}$, $c_{2}=2.3695\times10^{-10}$ J\,m$^{-1}$, and $c_{3}=-1.8229\times10^{-10}$ J\,m$^{-1}$.
\begin{figure}[ptbh]
\begin{centering}
\includegraphics[width=0.48\textwidth]{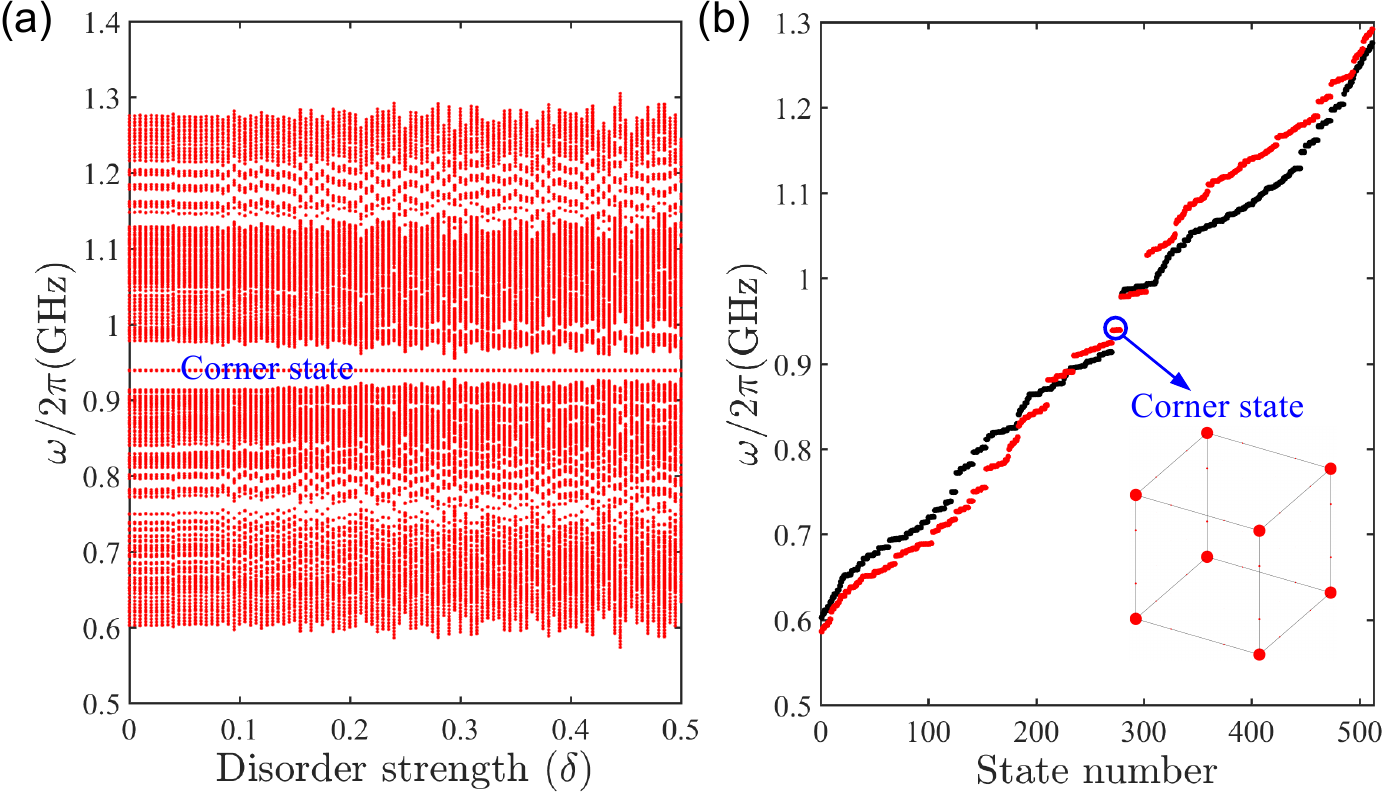}
\par\end{centering}
\caption{(a) Eigenfrequencies of the three-dimensional vortex lattice under different disorder strengths. (b) Eigenfrequencies of the three-dimensional vortex lattice in the absence of defects (black balls) and in the presence of defects (red balls). The blue circle indicates the topologically stable corner state with the inset showing the corresponding spatial distribution.}
\label{Figure7}
\end{figure}    
\section*{APPENDIX B: THE ROBUSTNESS OF THIRD-ORDER TOPOLOGICAL EDGE STATES}
To verify the topological nature of the corner states emerging in Fig. \ref{Figure3}, we calculate the eigenfrequencies of the three-dimensional vortex lattice with disorder and defects. The numerical results are presented in Figs. \ref{Figure7}(a) and \ref{Figure7}(b). Here, the disorder is introduced to all vortices by assuming the interlayer coupling parameter $\mu$ undergoing a random variation: $\mu\rightarrow\mu(1+\delta Z)$, where $\delta$ is the strength of disorder and $Z$ is a uniformly distributed random number between $-1$ and $1$. The defects are introduced to all vortices by assuming a shift to the coupling parameters $\zeta$, $\xi$, and $\mu$: $\zeta\rightarrow0.8\zeta$, $\xi\rightarrow0.8\xi$, and $\mu\rightarrow1.2\mu$. From Fig. \ref{Figure7}(a), one can clearly see that with the increasing of disorder strength, the corner state is very robust. Meanwhile, when defects are included, the frequency of corner states is perfectly pinned to 0.939 GHz, while the frequencies of hinge, surface, and bulk states are significantly modified [see Fig. \ref{Figure7}(b)]. These results thus support the conclusion that the third-order corner states emerging in our system are topologically stable.    

\end{document}